\documentclass[sigconf, natbib=true]{acmart}

\usepackage{hyperref}
\usepackage{multirow}
\usepackage{subcaption}
\usepackage{caption}
\usepackage{setspace}
\usepackage{amsmath}
\usepackage[english]{babel}
\definecolor{mygray}{gray}{0.5}
\usepackage{float}
\usepackage{flushend}
\usepackage{mathtools}
\usepackage{soul}
\usepackage{arydshln}
\usepackage{algpseudocode}
\usepackage{algorithm}
\usepackage{url}
\usepackage{xcolor}

\usepackage{graphicx}

\usepackage{enumitem}

\long\def\comment#1{}
\long\def\comments#1{}

\author{Parker Carlson}
\affiliation{%
  \institution{University of California, Santa Barbara}
  \city{Santa Barbara}
  \state{California}
  \postcode{93106}
  \country{USA}
}
\author{Wentai Xie}
\affiliation{%
  \institution{University of California, Santa Barbara}
  \city{Santa Barbara}
  \state{California}
  \postcode{93106}
  \country{USA}
}
\author{Rohil Shah}
\affiliation{%
  \institution{University of California, Santa Barbara}
  \city{Santa Barbara}
  \state{California}
  \postcode{93106}
  \country{USA}
}
\author{Tao Yang}
\affiliation{%
  \institution{University of California, Santa Barbara}
  \city{Santa Barbara}
  \state{California}
  \postcode{93106}
  \country{USA}
}
\renewcommand{\shortauthors}{Parker Carlson, Wentai Xie, Rohil Shah, and Tao Yang}

\begin{document}

\title{
Efficiency Optimizations for Superblock-based Sparse Retrieval 
} 

\begin{abstract}

\comments{
This paper studies the optimization of  superblock pruning  during top-$k$ online document retrieval
for learned sparse representations.
The previous work considers  a  sparse index as a set of superblocks with a sequence of document blocks
and conducts superblock-level  pruning  to quickly skip visitation of  many blocks  during  index traversal.
This paper proposes a simple yet effective lightweight superblock pruning scheme to 
reduce the overhead of aggregated superblock score bound computation 
while preserving relevance competitiveness of retrieval results.
It also devises
a compact index structure to facilitate fast superblock pruning,
and a simple zero-shot configuration.
This paper provides experiments to validate effectiveness and efficiency of lightweight superblock pruning
with a simple parameter configuration applicable to multiple datasets.
It also discusses when this lightweight scheme can outperform the baselines for in-domain and out-of-domain 
search with MS MARCO and BEIR datasets.
}


  Learned sparse retrieval (LSR) is a popular method for first-stage retrieval
  because it combines the semantic matching of language models with efficient CPU-friendly algorithms.
  Previous work aggregates blocks into ``superblocks'' to quickly skip the visitation of blocks during query processing
  by using an advanced pruning heuristic.
  This paper proposes   a simple and effective superblock pruning scheme that reduces the overhead of superblock score computation
  while preserving competitive relevance. 
  It combines this scheme with a compact index structure and a robust zero-shot configuration
  that is effective across LSR models and multiple datasets.
  This paper  provides an analytical justification and evaluation on the MS MARCO and BEIR datasets,
  demonstrating that the proposed scheme  can be a strong  alternative for efficient sparse retrieval.

\end{abstract}

\maketitle

\section{Introduction}

Fast and effective retrieval is a critical component of large-scale search systems.
It is also important for retrieval-augmented generation 
which is gaining in popularity~\cite{2020NIPS-Lewis-RAG,2024KDD-RAGsurvey}.
Learned sparse retrieval (LSR)~\cite{Formal2021SPLADE, Lin2021unicoil, mallia2021learning, shen2023lexmae,zeng2025scalingsparsedenseretrieval,lassance2024spladev3} 
can be used in the first-stage of the above systems as it can run fast on a low-cost CPU-only platform 
while delivering competitive relevance compared to dense retrieval which typically requires an expensive GPU server~\cite{Santhanam2021ColBERTv2,  Wang2022SimLM,  Liu2022RetroMAE}.
LSR can also be combined with dense retrieval
to further boost relevance on a CPU platform~\cite{2022LinearInterpolationJimLin,kuzi2020leveraging, 2024SIGIR-YangCDFS}.

Given a learned or traditional sparse index,
a major optimization for fast online inference is to use
dynamic index pruning that skips a significant portion of the index during query processing.
The state-of-the-art LSR retrieval algorithms ASC, BMP, Seismic and SP
all use block- or cluster-based pruning strategies~\cite{2024EMNLP-ASC,mallia2024BMP,2024SIGIR-SparseApproximate,2025SIGIR-Carlson-SP}.
Block-based skipping~\cite{2013CandidateFiltering, 2021WSDMliveBlock, mallia2024BMP}
divides groups of documents into blocks to estimate the block-wise maximum rank score among its documents
and prunes such a block when a block-wise maximum score  is below a threshold.
Similar pruning techniques are used in cluster-based skipping~\cite{2022ACMTransAnytime,2024EMNLP-ASC}.
Recently SP extends the work of BMP~\cite{mallia2024BMP} and ASC~\cite{2024EMNLP-ASC}
with  a superblock-level pruning scheme that aggregates both superblock-level maximum and average rank scores of documents~\cite{2025SIGIR-Carlson-SP}. 
A superblock contains a set of document blocks, and superblock pruning in SP increases index-skipping opportunities 
while maintaining competitive relevance when threshold 
overestimation~\cite{2012SIGIR-SafeThreshold-Macdonald, 2013WSDM-SafeThreshold-Macdonald, 2017WSDM-DAAT-SAAT}  is used.

One weakness when using overestimation is that there exist erroneous cases where almost all documents are removed
without producing enough top-$k$ results.
The pruning safeness protection added by average score computation in SP is not sufficient to prevent the pruning of almost all superblocks in such cases. 
Another weakness is that there is considerable overhead in maintaining and computing the average rank scores in SP.
In addition, they did not address index compression, and thus the impact of accessing compact data structures for two-level 
pruning is not well understood.
To address these weaknesses, 
we propose a lightweight superblock pruning (LSP) scheme which guarantees visitation of a fixed number of superblocks
while removing the average-bound-based safeguard to reduce overhead when using small block and superblock sizes.
We also propose compact data structures that significantly reduce memory overhead with negligible effects to retrieval time.

We provide an analysis to justify the advantages of top superblock inclusion 
while obtaining competitive relevance.
This also allows us to recommend simple parameter configurations for LSP under  popular 
retrieval depths ($k=10$ and 1000) without grid search. This is highly desirable for zero-shot retrieval. 
We have considered three versions of superblock pruning from a lightweight guarantee to more complex pruning schemes,
and we evaluated these strategies with the MS MARCO and BEIR datasets to assess when the proposed methods can outperform baselines.

Our evaluation shows that the simple LSP version is 1.8-17x faster than SP, and 1.8-12x faster than BMP.
It is 2x faster than SeismicWave for zero-shot retrieval with BEIR at various depths $k$. 
It is up-to 4.8x faster than SeismicWave for MS MARCO with $k=1000$ while  30\% slower for $k=10$.
When a grid search of all parameters is possible, it is up-to 78\% faster than SeismicWave for $k= 1000$,
while  it is  up-to  2x slower than SeismicWave for $k=10$. 

\vspace{-3mm}
\section{Background and Related Work}
\label{sect:background}

{\bf Problem definition.}
Given query $q$ for searching a collection of $D$ text documents:   $ \{d_i\}^D_{i=1}$,
sparse retrieval computes the dot product similarity of a given query $q$ and a document $d$ as:
$RankScore(d)  =L(q) \cdot  L(d_i)$
where $L(.)$ is a sparse vector of weighted term tokens for a document or a query.
Term weights typically use the BM25 formula~\cite{Robertson2009BM25},
or are learned through a BERT or LLM based model~\cite{Mallia2021deepimpact, Lin2021unicoil,
Formal2021SPLADE, shen2023lexmae, zeng2025scalingsparsedenseretrieval}.
$L(.)$ may contain original  and/or  expanded terms.
Online inference with sparse retrieval historically uses an inverted index to assist fast index traversal.

{\bf Partial index traversal with threshold-driven pruning}.
Given a sparse index, a popular efficiency optimization to skip portions of the index during query processing is dynamic pruning.
This strategy  computes the upper bound rank score of a candidate document $d$,
referred to as $Bound(d)$, satisfying  $RankScore(d) \le Bound(d)$.
If $Bound(d) \leq  \theta$, where $\theta$ is the rank score threshold to be in the top-$k$ list,
this document can be safely skipped.
Examples of dynamic pruning methods include  WAND~\cite{WAND}, 
BMW~\cite{BMW}, VBMW~\cite{Mallia2017VBMW}, and  MaxScore~\cite{Turtle1995}. 
A  retrieval method is  called {\em rank-safe} if it  guarantees that the top-$k$ documents returned are the $k$ highest scoring documents.
All of the above algorithms are rank-safe.
Threshold over-estimation is  a ``rank-unsafe'' skipping strategy that  deliberately over-estimates the current top-$k$ threshold
by a factor~\cite{2012SIGIR-SafeThreshold-Macdonald, 2013WSDM-SafeThreshold-Macdonald, 2017WSDM-DAAT-SAAT}.

{\bf Block or cluster based sparse index pruning.}
The state-of-the-art fast sparse retrieval algorithms for learned sparse retrieval
ASC, BMP, Seismic, and SP~\cite{2024EMNLP-ASC,mallia2024BMP,2024SIGIR-SparseApproximate,2025SIGIR-Carlson-SP}
all use block- or cluster-based pruning strategies. 
There is a large body of studies on cluster-based document retrieval  in traditional IR
(e.g.  ~\cite{liu2004cluster,2022SIGIR-KurlandClusterRank,kurland2008opposite})
for re-ranking an initially retrieved list by creating clusters of similar  documents in this list.
Then the clusters are ranked, and the cluster ranking may be further transformed to a document ranking.
Block-based index skipping~\cite{2013CandidateFiltering, 2021WSDMliveBlock, mallia2024BMP}
divides  documents into blocks to estimate the block-wise maximum rank score for pruning.
Conceptually, cluster-based skipping follow a similar strategy~\cite{2004InfoJ-ClusterRetr,2017ECIR-SelectiveSearch, 2022ACMTransAnytime}.
BMP~\cite{mallia2024BMP} optimizes execution with score quantization, SIMD $BoundSum$ computation,
partial block sorting, and query term pruning.
A major optimization in Seismic~\cite{2024SIGIR-SparseApproximate}
is aggressive static inverted index pruning while fully scoring documents with an unpruned forward index.
Like BMP~\cite{mallia2024BMP}, Seismic also incorporates threshold overestimation, query term pruning,
and dynamic cluster (block) maximum pruning.
SeismicWave~\cite{2024CIKM-SeismicWave} extends Seismic to improve the relevance of approximate search by sorting blocks by their visitation heuristic
and adding a post processing stage to select relevant documents by using an extra data structure
with a  document-level proximity graph.
ASC~\cite{2024EMNLP-ASC}
and SP ~\cite{2025SIGIR-Carlson-SP} extend the above pruning studies and compute both maximum and average rank scores 
to  assure  probabilistic rank-safeness for   competitive relevance.
In terms of clustering or block assignment methods, the above studies  employ a similarity-based method to group similar documents together,
e.g. a bipartite partitioning algorithm~\cite{dhulipala2016bp, mackenzie2019bp} or k-means~\cite{chierichetti2007finding}.

{\bf Index compression.} 
Earlier IR studies have  proposed  various index compression techniques, e.g.~\cite{
moffat2018index, 
ottaviano2014partitioned,
anh2010index,
zhang2008performance,
trotman2016vacuo,
yan2009inverted,
2015Lemire},
and a study by Mallia et al.~\cite{mallia2019experimental}
comparing a number of these compression techniques found that SIMD-BP128~\cite{2015Lemire} provides 
a good trade-off
 between space reduction and decoding speed.
These techniques are  effective  for  traditional retrieval algorithms with inverted indices.
To accommodate recently-developed cluster- or block-based algorithms,  
new investigations are desired to ensure a proper tradeoff of space reduction  and query processing latency when incorporating with two-level pruning.
BMP~\cite{mallia2024BMP} by Mallia et al. has devised a compact scheme that works well with its block filtering strategy, 
while  Seismic, ASC,  and SP  do not address index compression. 
This paper assesses the use of BMP compression for LSP and 
designs an improved  scheme that   works effectively with the proposed method, based on SIMD-BP and
an integer list packing method previously proposed for GPU encoding~\cite{2019CIKM-GPUdecode-Mallia}.

{\bf Other orthogonal efficiency techniques.}
There are orthogonal techniques to accelerate learned sparse retrieval through
 BM25-guided search, model training,  
and index re-organization,
e.g. ~\cite{mallia2022faster,20232GT,
lassance2022efficiency,
mackenzie2022accelerating}.
Static index pruning for SPLADE is studied in ~\cite{2023SIGIR-Qiao,2023SIGIR-SPLADE-pruning}
and is employed extensively by Seismic~\cite{2024SIGIR-SparseApproximate}.
Our work is complementary to these studies, and can benefit from the use of these techniques.

\section{Block/Superblock Pruning and Design Consideration}
\label{sect:design}

First, we describe block-based filtering in BMP and superblock pruning in SP. 
We assume that a document collection is divided into $N$ blocks.
Following~\cite{mallia2024BMP, 2025SIGIR-Carlson-SP},
these blocks are formed based on  similarity, and  each block uniformly contains $b$ documents.
During online retrieval, BMP visits these blocks in a decreasing order of their $BoundSum$ values, where
$BoundSum(B)$ is a maximum rank score bound of documents within block $B$. 
Let $\theta$ be the estimated~\cite{mallia2020comparison} or currently known top-$k$ rank score of documents scored so far by this retriever for a query. 
The visitation to block $B$ can be pruned if $ BoundSum(B) \leq \theta $.
The block-level rank score bound is estimated using maximum term weights within each block.
This bound can be very loose, especially when a cluster contains documents with disparate rank scores.
Because of loose bound estimation, blocks composed of low-scoring documents often do not satisfy the condition $ BoundSum(B) \leq \theta $
and are not pruned.
Threshold overestimation~\cite{2012SIGIR-SafeThreshold-Macdonald, 2013WSDM-SafeThreshold-Macdonald, 2017WSDM-DAAT-SAAT} with
$\mu$-approximation addresses that
by changing  the above pruning condition as  $BoundSum(C_i) \leq \frac{\theta}{\mu} $
where $0< \mu <1$.
However, this change  can incorrectly prune desired relevant documents when their cluster bound estimation is tight.

To identify low-scoring blocks earlier and 
address the above weakness, SP furthers uniformly aggregate a sequence of $c$ consecutive document blocks into one superblock, 
and online inference  traverses the index in a top-down manner.
Namely, SP visits all superblocks first and then visits the blocks of unpruned superblocks.
Finally, it visits and scores documents within unpruned blocks.

For each superblock $X$, SP computes the rank score bound ($SBMax(X)$)
and average score bound ($\overline{SBMax}(X)$)
of documents within superblock $X$ as follows 
\begin{equation}
\label{eq:subtreemax} 
SBMax(X) = \sum_{t \in Q}   q_t\cdot W_{X,t};  \
\overline{SBMax}(X) =
\sum_{t \in Q} q_t \cdot \overline{W}_{X,t}
\end{equation}
where $W_{X,t}$ and $\overline{W}_{X,t}$ are  the maximum and average weight of term $t$ among documents of superblock $X$,
respectively.

Then SP adds an extra pruning safeness guard by using the average score bound.
Any superblock $X$ is pruned when its maximum and average superblock bounds satisfy
\begin{equation}
\label{eq:prune1}
   SBMax(X)  \le \frac{\theta}{\mu}
\end{equation}
and  
\begin{equation}
\label{eq:prune2}
   \overline{SBMax}(X) \le \frac{\theta}{\eta}
\end{equation}
where parameters $\mu$ and $\eta$ satisfy $0<\mu \leq \eta \le 1$. 
This two-level pruning of SP is similar to the $(\mu,\eta)$-approximation
in  ASC~\cite{2024EMNLP-ASC}, except SP incorporates additional efficiency optimizations.
The use of Inequality~(\ref{eq:prune1}) retains $\mu$ overestimation safeness guarantees,
while Inequality~(\ref{eq:prune2}) ensures probabilistic safeness.

\noindent
{\bf Design Considerations.}
Our considerations to accelerate SP are:

\begin{itemize}[leftmargin=*]
\item {\bf Visit fewer superblocks.}
To accelerate retrieval in a block- and superblock-based index structure,
we seek more opportunities to prune superblocks.
As superblock maximum bounds are generally loose,
selecting a relatively smaller $\mu$ value exposes
opportunities for filtering them out using Inequality~(\ref{eq:prune1}), guarded by Inequality~(\ref{eq:prune2}).  
Define the bound tightness of superblock $X$ as $max_{d\in X} RankScore(d)$
divided by $SBMax(X)$.  
Figure~\ref{fig:superblock_tightness} depicts the distribution of the average superblock bound tightness on 
the MS MARCO Dev queries with block size $b=8$  and with $c=16$ blocks per superblock.
The tightness value varies from about 0.2 to 1.
To prune most superblocks, $\mu$ would need to be set as low as possible, near 0.2.

\begin{figure}
   \includegraphics[width=\columnwidth]{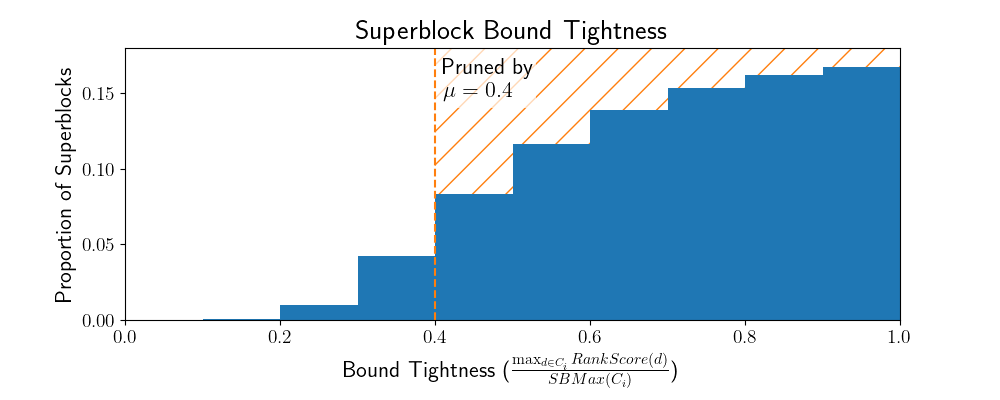}
   \caption{ Distribution of bound tightness values of superblocks on MS MARCO Dev set with b=8, c=16}
   \label{fig:superblock_tightness}
\end{figure}

\item {\bf Erroneous pruning with $\mu$-overestimation}. 
There are cases when $\mu$ is small where Inequality~(\ref{eq:prune1}) is true for any superblock $X$ in a dataset while 
Inequality~(\ref{eq:prune2}) is also true for most superblocks even when $\eta=1$.
Then, Inequalities (\ref{eq:prune1}) and (\ref{eq:prune2}) erroneously prune enough superblocks that
there are not enough documents left to contribute $k$ results.
Figure~\ref{fig:erronous} shows how often erroneous pruning occurs in processing  MS MARCO 
Dev queries with SPLADE by SP with retrieval depth $k=1000$ when $\mu \leq 0.5$ and $\eta=1$.  
The $y$-axis shows the percentage of queries failed to produce any result in orange (big dashes),
the percentage of queries producing results less than $k$ in green (small dashes), and the percentage of queries producing exactly top $k$ results in blue (solid),
when  $\mu$ varies from 0.1 to 0.5 on the $x$ axis. 
When $\mu \ge 0.5$, the above erroneous pruning
does not occur for these queries with SPLADE,  but it happens with E-SPLADE as shown in Table~\ref{tab:main}
of Section~\ref{sect:eval}.  

\begin{figure}
   \includegraphics[width=\columnwidth]{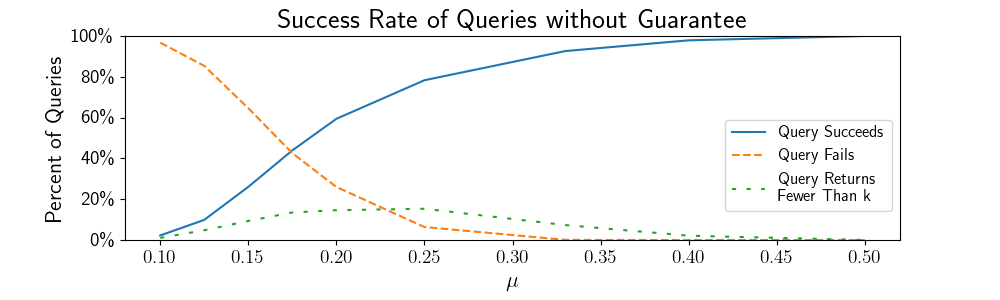}
   \caption{Successful, partially successful, and failed  top-$1000$ query processing for MS MARCO Dev set when $\mu$ varies}
   \label{fig:erronous}
   \vspace{-2mm}
\end{figure}

\item {\bf Significant  overhead to maintain and compute average rank score bounds}. 
The additional storage and computation required to find $ \overline{SBMax}(X) $ for each superblock
sometimes outweighs the benefit of additional safeness when there is a large number of superblocks.
\item  {\bf Index compression.}
The current SP implementation stores block- and superblock-bounds uncompressed.
Because computing these bounds is latency-sensitive, 
it is desirable to design a proper compact data structure that works well with superblock pruning.

\end{itemize}

Our design pursues the following strategies:

\begin{itemize}[leftmargin=*]
\item 
To avoid the erroneous  pruning, our idea is  to  impose
an extra condition guaranteeing top-$\gamma$ superblocks with safe pruning 
during index traversal. More details are in 
Section ~\ref{sect:topinclusion}
with a design justification in Section~\ref{sect:analysis}.
\item We view the average maximum bound condition as an extra guard to work in tandem  with the $\mu$-based pruning for improved
safeness. Our finding is that with inclusion of top-$\gamma$ superblocks,
the average maximum bound check with Inequality~(\ref{eq:prune2})
may not be necessary, as shown in Section~\ref{sect:eval},
and it can be removed for improved time and space efficiency.

\item We devise a compact data structure for superblocks and blocks, that adds reasonable access overhead when conducting superblock
pruning. More details are in Section~\ref{sect:compress}. 
\end{itemize}

\section{Lightweight  Superblock Pruning}\label{sect:motivation}

\subsection{Pruning with top-$\gamma$ superblock inclusion}
\label{sect:topinclusion}

\begin{figure}
   \includegraphics[width=0.95\columnwidth]{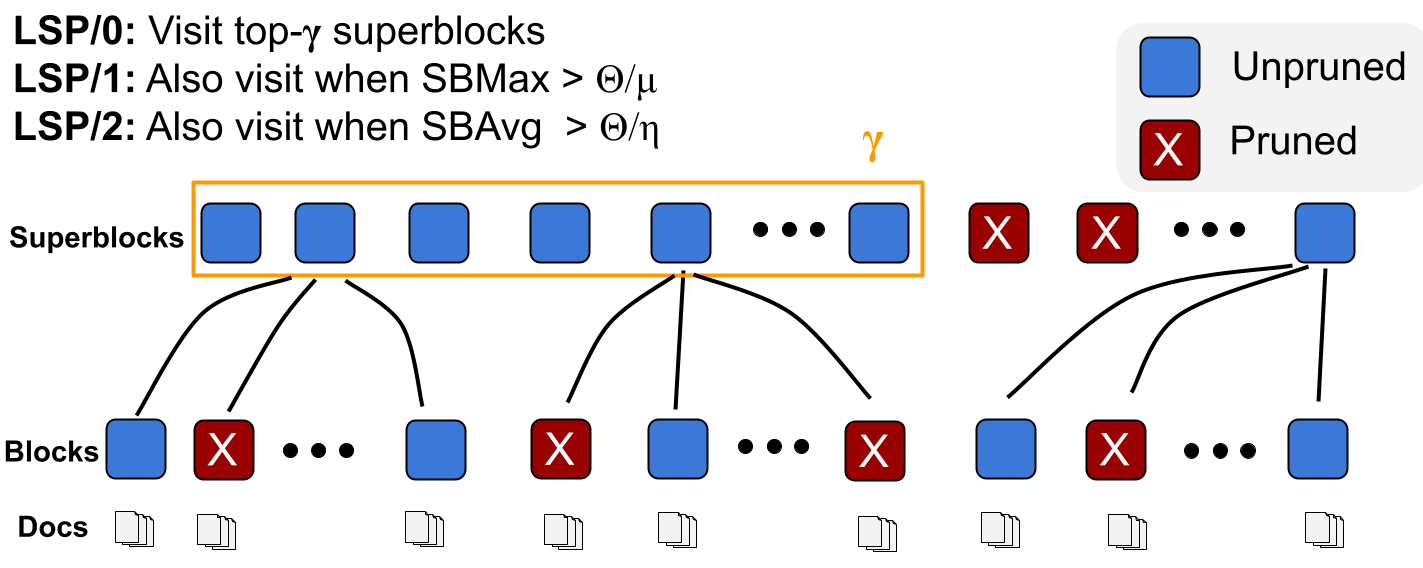}
   \caption{Traversal with superblock pruning in LSP}
   \label{fig:tree}
   \vspace{-2mm}
\end{figure}

Our idea to address erroneous pruning 
is to impose a simple search guarantee that retrieval  always searches
the top-$\gamma$ superblocks with safe pruning. 
That comes from an observation that superblocks with the largest maximum bound values have
a high chance to deliver the top-$k$ relevant results, even though the estimation of maximum score bound can be fairly loose.
There is no need to attempt to prune such superblocks.
To justify the above design with a proper $\gamma$, Section~\ref{sect:analysis} provides
 a statistical analysis. 

Imposing the guaranteed inclusion of top-$\gamma$ superblocks may sufficiently ensure that
top-$k$ documents will be found with a high confidence when $\gamma$ is properly chosen. 
That means this safeness protection has a large overlap with
the role of Inequality (\ref{eq:prune1}) and/or  Inequality (\ref{eq:prune2}) in SP.
We consider three versions to simplify SP while avoiding erroneous pruning,
as shown in Figure~\ref{fig:tree}:

{\bf Version 
LSP/0:}  Lightweight superblock pruning with guaranteed top-$\gamma$ superblock inclusion. 

\noindent
\fbox{%
\begin{minipage}{0.45\textwidth}
\begin{itemize}[leftmargin=*]
	\item  Identify the top-$\gamma$ superblocks with the highest $SBMax(X)$ values
as long as their superblock level rank bound satisfies $SBMax(X)  \ge   \theta$.
	\item  Visit the blocks of the above selected superblocks and prune according to $\frac{\Theta}{\eta}$, following SP, and score the documents within these unpruned
blocks. 
\end{itemize}
       
\end{minipage} 
}

This is a simple version of block-based retrieval with superblock-level pruning without
using Inequalities (\ref{eq:prune1}) or (\ref{eq:prune2}).
The block-level pruning and visitation behave the same as BMP and SP with threshold overestimation applied using $\eta$ (typically 1.0).
Compared to BMP, the main advantage is that superblock-level filtering avoids  the visitation of large  blocks of documents 
without accessing their block-level or document-level information.  

{\bf Version LSP/1}:  Lightweight superblock pruning with guaranteed top-$\gamma$ superblock inclusion and $\mu$-overestimation.

\noindent
\fbox{
\begin{minipage}{0.45\textwidth}
\begin{itemize}[leftmargin=*]
\item  This is the same as LSP/0 except that it includes  additional superblocks $X$ satisfying $SBMax(X)  > \frac{\theta}{\mu}$.
\end{itemize}
\end{minipage} 
}

This version ensures that top-$\gamma$ superblocks are searched,
then adds the additional safeness of threshold overestimation.
Thus, the average rank score of any top-$k$ results produced by this algorithm is 
competitive to that of rank-safe retrieval within a factor of $\mu$~\cite{2024EMNLP-ASC}.

{\bf Version LSP/2:}
 Superblock pruning  with guaranteed $\gamma$-superblock inclusion and $(\mu-\eta)$-pruning.

\noindent
\fbox{%
\begin{minipage}{0.45\textwidth}
\begin{itemize}[leftmargin=*]
	\item  This algorithm first  guarantees top $\gamma$ superblocks $X$ with safe pruning as long as $SBMax(X)  \ge   \theta$.
	\item  Then it behaves the same as SP for other superblocks.
\end{itemize}
\end{minipage} 
}

This version enhances the safeness of LSP/1
by adding probabilistic safeness with $\eta$ as described in ~\cite{2024EMNLP-ASC,2025SIGIR-Carlson-SP}. 
Our finding is that this extra safeness is largely overlapped with top-$\gamma$ superblock inclusion.
It is less valuable when the superblock size $b\times c$ is relatively small in which the average superblock rank bound 
is not too far away from the maximum.

\subsection{Choice of $\gamma$ for top superblock inclusion}
\label{sect:analysis}

This subsection analyzes suitable $\gamma$ values for the inclusion of top-$\gamma$ superblocks.
Given $N$ superblocks $S_1, \cdots, S_N$ sorted   by their rank score bound $SBMax$ values in a non-ascending order,
our goal is to  find a value $\gamma$   
so that with a high confidence superblock $S_i$  $(\gamma+1 \leq  i \le N)$ does  not contain  a document that  appears in
the  top-$k$ document list  of a safe pruning algorithm.

For superblock $S$,  we define an $SBMax$ ratio value as $\frac{SBMax(S)}{\max(SBMax(S))}$, representing
its closeness to the top-1 superblock $SBMax$ value for a query. 
From a training dataset, we can approximately derive a distribution for the $SBMax$ ratio values
that superblocks can take.

Now let $X$ be a random variable that takes a value between 0 and 1, following the above $SBMax$ ratio distribution.
We consider a random sample of size $N$ from this distribution, denoted by $X_1, X_2, \cdots, X_N$.
A maximum order statistic is the result of arranging these random variables following our retrieval  algorithm in a
non-increasing order. Namely $X_{(1)} \ge X_{(2)} \ge \cdots \ge X_{(N)}$.

The standard formula for cumulative probability $P(X_{(i)} \le x)$ CDF of the $i$-th largest order statistic  is: 
\[
P(X_{(i)} \le x) = \sum_{j=N-i+1}^{N} \binom{N}{j} [P(X \le x)]^j [1 - P(X \le x)]^{N-j}. 
\]
$P(X \le x)$  is obtained using a training dataset
by computing the distribution of each possible $SBMax$ value over all superblocks for different training queries.
The formula of $P(X_{(i)} <x)$  is similar.

We also divide all possible $SBMax$ values into a set of bins $B_j$.
Let $R$ be an event that a superblock contains a top-$k$ document. 
Let $I$ be an event that a superblock does not contain a top-$k$ document. 
Let $P_{\gamma} (I)$ be the confidence in probability that the $\gamma$-th superblock $S_{\gamma}$ does not contain a top-$k$ document.  Then,

$$ P_{\gamma}(I) = 1- P_{\gamma}(R) = 1- \int_{0}^1 P(R | x) P_{\gamma}(x) dx \approx 1- \sum_{j} P(R| B_j) \cdot P_{\gamma}( B_j). 
$$
\comments{
where $P_{\gamma}(R| B_j) = P(R| B_j)$ because that is independent of the superblock position.
}

Here $P_{\gamma}(B_j)$ is the probability of a specific bin that the value of $X_{(\gamma)}$ can belong to, and  it
is determined by the difference between the cumulative probabilities at its right and left boundaries:

\vspace{-2mm}
$$P_{\gamma}(B_j) = P(X_{(r)} \le x_{j,r}) - P(X_{(r)} < x_{j,l})  $$
where:
\begin{itemize}
    \item $x_{j,r}$ is the right boundary  SBMax ratio value of this  bin $B_j$;
    \item $x_{j, l}$ is the left boundary  SBMax ratio value of this  bin.
\end{itemize}

For example  with two bins: $[0, 0.5)$ and $[0.5, 1]$,  $P_{\gamma}(B_2) = P(X_{(\gamma)} \le 1) - P(X_{(\gamma)} < 0.5)$.

\begin{figure}
   \includegraphics[width=\columnwidth]{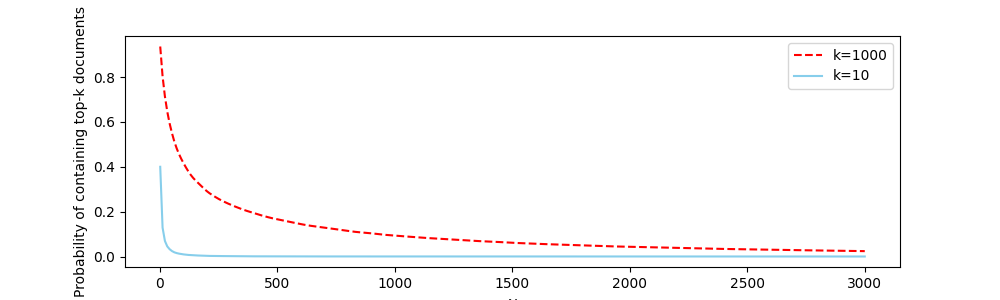}
   \caption{Probability of containing top-$k$ documents at different sorted top superblock positions with $1\leq \gamma \leq 3000$} 
   \label{fig:waterfall}
\end{figure}

Figure~\ref{fig:waterfall} shows the $P_{\gamma}(R)$ value distribution  when 
$\gamma$ varies from 1 to 3,000 for $k=10$ and $k=1000$, derived from 10,000 MS MARCO Passage training queries,
 and each superblock contains  128 documents  (e.g. $b=8$ and $c=16$).
The curves for $b=4$ and $16$ are very similar. Table~\ref{tab:predict} shows $P_{\gamma}(R)$ at selected values of $\gamma$, $k$, and $b$.
The main takeaways are: 
\begin{itemize}[leftmargin=*]
\item $P_{\gamma}(R)  > P_{\gamma+1}(R)$.  The confidence to find a top-$k$ document in superblock $S_{\gamma}$
decreases monotonically as $\gamma $ becomes large.
\item  With $k=10$, 
confidences  among different block/superblock size are about the same for fixed $\gamma$ in Table~\ref{tab:predict}. 
Our evaluation  chooses $\gamma=250$ and $500$ with high 99.6+\% confidence. 

\item With $k=1000$ and same $\gamma$ value, 
confidences  among different block/superblock sizes differ by about or within 1\% when $\gamma \ge 1000$. 
Our evaluation  chooses $\gamma=1000$ and $2000$ with above 90\% and 95\% confidence. 
We choose these confidence levels because the above analysis is  a conservative estimation of  $\gamma$ as it 
does not consider the impact of similarity-based clustering in block construction.
Once a superblock with a high $SBMax$ value contains one top-$k$ document,
there is a good chance that it contains more than one such documents for large values of $k$. 
 
\end{itemize}

\begin{table}[thbp] 
    \centering 
    \caption{Confidence $P_{\gamma}(I)$ based on MS MARCO training data }
    \small
    \label{tab:predict}
    \resizebox{\columnwidth}{!}{ 
        \begin{tabular}{l|cccccc}
        \hline 
        {\bf  $\gamma$  } &{100} & {200} & {300} & {1000} & {2000} & {3000} \\
        \hline 
         $k=10, b\times c=64$ & 98.9\% & 99.6\% & 99.8\% & $\sim$ 100\% & $\sim$100\% & $\sim$100\% \\
         $k=10, b\times c=128$ & 99.0\% & 99.6\% & 99.8\% & $\sim$100\% & $\sim$100\% & $\sim$100\% \\
         $k=10, b \times c=256$ & 99.0\% & 99.6\% & 99.8\% & $\sim$100\% & $\sim$100\% & $\sim$100\% \\
        \hline
        \hline 
         $k=1,000, b\times c=64$ & 56.4\% & 69.4\% & 74.6\% & 89.4\% & 95.0\% & 97.1\% \\
         $k=1,000, b\times c=128$ & 59.6\% & 71.4\% & 77.3\% & 90.8\% & 95.8\% & 97.6\% \\
         $k=1,000, b \times c=256$ & 62.2\% & 73.7\% & 79.4\% & 91.9\% & 96.5\% & 98.1\% \\
        \hline
        \end{tabular}
        }
\end{table} 

\subsection{Index compression}\label{sect:method_comp}
\label{sect:compress}

{\bf Compact data structure for block and superblock maximum weights}.
BMP supports the option of storing their block bounds in an uncompressed dense format and in a sparse format
(denoted BMP-Dense and BMP-Sparse).
The sparse format yields 5.5GB for MS MARCO with
$b$=8 while the dense format yields 30GB. 
However, the use of BMP-Sparse for LSP is
not effective for random access required by superblock pruning. 
For example, for safe search using  BMP-Sparse is up to 76x slower
as shown in Section~\ref{sect:eval_comp}.

We seek to adopt SIMDBP for superblock and block level maximum weights because   
SIMDBP-128 is found to be highly effective for traditional 
inverted indices~\cite{2015Lemire,mallia2019experimental}.
SIMPDBP-128 organizes a list of integers to store in the following format
\fbox{$S_0$} \fbox{$G_{0,0}, \cdots, G_{0,127}$} \fbox{$S_1$} \fbox{ $G_{1,0} \cdots G_{1,127}$}, $\cdots$.  
Each group $G_i$ contains 128 integers using  at most $w$ bits where $w \leq 32$ and these integers  are stored 
in 128*$w$ bits with a compact form. When decoding, these integers are recovered into 32-bit integers. 
Group $S_i$ is called a selector group which contains 128 numbers representing the maximum width
for each of the following groups: $G_{i,0}, \cdots, G_{i,127}$.
  
\begin{figure}
   \includegraphics[width=0.8\columnwidth]{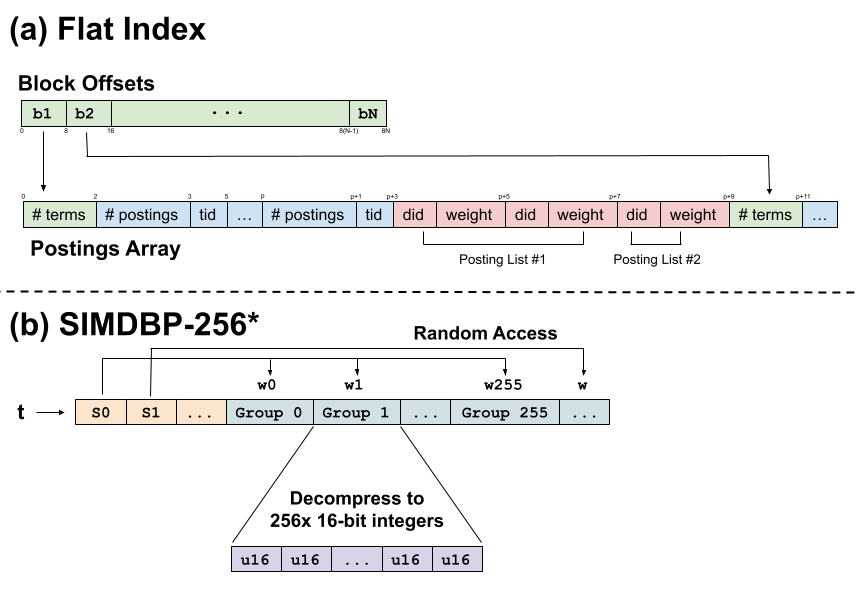}
\caption{(a) Flat inverted index structure for documents within a block.
(b) SIMDBP-256* data structure layout for 
storing maximum term weights for each block or superblock } 
\vspace{-5mm}
   \label{fig:combo_diagram}
\end{figure}

We design a customized version of SIMDBP with 256 integers per group optimized for storing term maximum weights 
of a superblock or a block, 
   as shown in Figure~\ref{fig:combo_diagram}(b).
\vspace{-2mm}
\begin{itemize} [leftmargin=*]
    \item We let each group store 256 block-level or superblock-level term maximum weights consecutively for the same level in an index forest structure.
Notice that a 16-bit integer is sufficient to   match the width limit of $BoundSum$  and $SBMax$ accumulation registers, and thus
we uncompress a group of  data to 256 16-bit integers instead of 128 32-bit integers used in SIMDBP-128.
The use of 16-bit integers doubles the number of integers processed per SIMD instruction for faster decoding.
    \item For every 256 groups, the original SIMDBP method  stores a group of "selectors" specifying  the bit-width of the following 256 groups.
    This is effective for decoding of these integers with a sequential access.
    Instead, we store the selectors at the start of each list of maximum term weights 
    which allows us to efficiently decompress any individual group in a random position.
\end{itemize}
\vspace{-2mm}

The second idea is similar to a layout by Mallia et al.~\cite{2019CIKM-GPUdecode-Mallia} for GPU decoding.
The main advantage in our context is that moving selectors in advance allows the quick access of any block at  a random position.
Because of superblock-level pruning, some blocks may be accessed in a non-consecutive manner, causing random access. 
We denote the above customized design  as SIMDBP-256*,
and our evaluations show that it is up to 1.5x faster than  SIMDBP-256 by enabling random access to block maximum weights. 

\noindent
{\bf Quantization for block/superblock maximum weights.}
BMP uses 8-bit quantization to store the block maximum weights of terms.
We find that the use of 4-bit quantization is sufficient for both block and superblock maximum weights.
The benefit of 4-bit quantization is a large space reduction of block and superblock maximum weights.
It also accelerates decompression and thus maximum rank score computation
while retaining competitive relevance, as shown in our evaluation. 
It should be noted that we still follow BMP to use 8-bit quantization for document-level term weights.

\noindent
{\bf Document index}.
We consider four methods to store documents.
We consider each method as a compact data structure,
and do not fully evaluate applying lossless compression techniques. 

\noindent
{\bf 1. BMP-Inv.}
BMP uses the terminology block forward index for its document index,
meaning it stores an inverted index data structure within each block.
We can use this methodology as an uncompressed document index (which we denote BMP-Inv for clarity).

This index is implemented using a Rust nested vector structure to store the term weights of each document.  
Like C++, each vector in Rust stores a pointer, a length, and a capacity,
each vector requires a significant overhead of 24 bytes on a 64-bit system.
For 8.8M MS MACRO passages with b=8, this means the vectors themselves require 15.5 GB when loading data to memory, and only about 3 GB is used to store the postings.
Notice that when data is stored on disk, pointers and capacity per vector are not needed.
BMP and SP both report the size of their indexes as the size on-disk.

\noindent
{\bf 2. Compact-Inv.}
We consider an optimized version of BMP-Inv.
We limit block sizes to 256 documents per block, and thus can use a single byte to represent the length.
We limit the index to 65k unique terms and store term-ids as two bytes.
Additionally, at query time when the index is fixed, the length and capacity of all vectors are identical
and we remove this redundancy.

\noindent
{\bf 3. Flat-Inv.}
We reorganize the data of Compact-Inv to
a single array instead of using nested Rust vector representations,
as shown in Figure~\ref{fig:combo_diagram}(a).
Additionally, we store a second list containing offsets to each block in the consolidated postings array.
We have also tried applying SIMDBP-256* and Simple8b~\cite{anh2010index} to the postings array after using delta encoding on term IDs and block offsets,
following findings of  ~\cite{mallia2019experimental}.
However, neither method significantly reduced the size of our flat data structure.
Therefore we opt to keep element values of the compact postings array uncompressed.

\noindent
{\bf 4. Fwd.}
We also adopt a simple forward index design from  Seismic~\cite{2024SIGIR-SparseApproximate,2024CIKM-SeismicWave},
which stores each document independently. 
Also following Seismic, we score documents using the entire query and only use the pruned query for candidate generation.
We apply this technique to all four document scoring methodologies.

\begin{table*}[thbp]
    \centering
    \caption{ MS MARCO passages:  mean response time ($ms$) and relevance of LSP/0 and the baslines with  two fixed configurations}
    \small
    \label{tab:main}
    \resizebox{2.0\columnwidth}{!}{
        \begin{tabular}{l|ccc|ccc|cc|cc|c|ccc|ccc|c}
        \hline
        \hline
        & \multicolumn{11}{c|}{{\bf SPLADE} (In-Domain Parameters)} & \multicolumn{7}{c}{{\bf E-SPLADE} (Zero-Shot Parameters)} \\ 
        \hline
        {\bf Config} & \multicolumn{3}{|c|}{{\bf Config. 1}} & \multicolumn{3}{|c|}{{\bf Config. 2}} & \multicolumn{2}{|c|}{{\bf DL19}} & \multicolumn{2}{|c|}{{\bf DL20}} & {\bf Size} & \multicolumn{3}{|c|}{{\bf Config. 1}} & \multicolumn{3}{|c|}{{\bf Config. 2}} & {\bf Size} \\
         & MRR & Re & MRT & MRR & Re & MRT & nDCG & Re & nDCG & Re & {\bf (GB)} & MRR & Re & MRT &  MRR & Re & MRT &  {\bf (GB)} \\
        \hline
        {\bf Method} & \multicolumn{11}{c|}{$k$=10} & \multicolumn{7}{|c}{$k$=10}\\  
        \hline
        & \multicolumn{11}{c|}{{\bf Uncompressed}} & \multicolumn{7}{|c}{{\bf Uncompressed}} \\  
        BMP & 38.20 & {\bf 66.97} & 2.53 & 38.11 & {\bf 66.99} & 3.18 & 73.16 & 17.25 & 71.97 & 24.54 & 19.2 & 38.69 & {\bf 66.98} & 0.629 & 38.81 & 67.44 & 0.744 & 24.5\\
        SP & 37.28 & 64.02 & 1.40 & 38.08 & 66.92 & 2.09 & 73.16 & 17.25 & 71.97 & 24.54 & 30.8 & \multicolumn{6}{c|}{Fail due to erroneous pruning}  & 37.0 \\ 
        Seismic-Wave & {\bf 38.25} & 66.70 & {\bf 0.297} & 38.27 & 66.89 & {\bf 0.403} & {\bf 73.28} & {\bf 17.34} & {\bf 71.99} & 24.47 & 7.79 & 38.78 & 67.42 & 0.766 & 38.79 & 67.40 & 0.809 & 11.2 \\
        LSP/0 & 38.14 & 66.42 & 0.425 & {\bf 38.28} & 66.88 & 0.649 & 73.13 & 17.25 & 71.84 & 24.46 & 18.8 & {\bf 38.91} & 67.35 & 0.402 & {\bf 38.92} & {\bf 67.59} & 0.506 & 20.8\\
        & \multicolumn{11}{c|}{{\bf Compressed}} & \multicolumn{7}{|c}{{\bf Compressed}}\\  
        MaxScore & -- & -- & -- & 38.11 &  {\bf 66.99} & 75.7 &  73.16 &  17.25 & 71.97 & {\bf 24.54} & {\bf 1.74} & -- & -- & -- & 38.81 & 67.44 & 8.06 & {\bf 2.13}  \\
        BMP & 38.17 & {\bf 66.97} & 4.20 & 38.11 & {\bf 66.99} & 5.41 & 73.16 & 17.25 & 71.97 & {\bf 24.54} & 17.4 & 38.67 & 66.93 & 1.11 & 38.81 & 67.44 & 1.28 & 24.5\\
        LSP/0 & 38.14 & 66.44 & 0.501 & {\bf 38.28} & 66.89 & 0.878 & 73.13 & 17.25 & 71.84 & 24.46 & 8.99 & {\bf 38.91} & 67.35 & 0.455 & {\bf 38.92} & {\bf 67.59} & 0.570 & 12.1 \\
        + 4-bit Quant. & 38.08 & 66.36 & 0.347 & 38.23 & 66.84 & 0.562 & 73.13 & 17.25 & 71.93 & 24.50 & 5.52 & 38.86 & 67.10 & {\bf 0.327} & 38.89 & 67.35 & {\bf 0.424} & 7.65  \\

        \hline
        & \multicolumn{11}{c|}{$k$=1000} & \multicolumn{7}{|c}{$k$=1000}\\
        \hline
        & \multicolumn{11}{c|}{{\bf Uncompressed}} & \multicolumn{7}{|c}{{\bf Uncompressed}}\\  
        BMP & 38.20 & {\bf 98.33} & 11.2 & 38.11 & {\bf 98.36} & 13.9 & {\bf 73.16} & 82.91 & 71.97 & {\bf 83.91} & 47.6 & 38.69 & {\bf 97.94} & 14.3 & 38.81 & {\bf 98.03} & 15.5 & 58.2 \\
        SP & 38.09 & 98.24 & 5.70 & 38.09 & 98.32 & 11.1 & {\bf 73.16} & 82.95 & 71.97 & {\bf 83.91} & 49.0 & 34.52 & 96.48 & 25.8 & 34.52 & 96.64 & 30.1 & 58.3 \\
        Seismic-Wave & {\bf 38.30} & 98.17 & 4.42 & {\bf 38.30} & 98.26 & 4.97 & 73.14 & {\bf 82.97} & {\bf 71.99} & 83.81 & 11.3 & 38.80 & 97.85 & 8.09 & 38.80 & 98.00 & 8.57 & 16.4 \\
        LSP/0 & 38.29 & 98.10 & 1.62 & 38.29 & 98.34 & 3.08 & 73.12 & 82.32 & 31.86 & 83.09 & 65.7 & {\bf 38.91} & 97.90 & 2.01 & {\bf 38.91} & {\bf 98.03} & 2.35 & 69.1 \\
        & \multicolumn{11}{c|}{{\bf Compressed}} & \multicolumn{7}{|c}{{\bf Compressed}}\\ 
        MaxScore & -- & -- & -- & 38.11 & {\bf 98.36} & 124 & {\bf 73.16} & 82.91 & 71.97 & {\bf 83.91} & {\bf 1.74} & -- & -- & -- & 38.81 & {\bf 98.03} & 13.2 & {\bf 2.13} \\
        BMP & 38.18 & 98.32 & 11.9 & 38.11 & {\bf 98.36} & 14.8 & {\bf 73.16} & 82.91 & 71.97 & {\bf 83.91} & 29.0 & 38.68 & 97.92 & 7.64 & 38.81 & {\bf 98.03} & 8.86 & 41.2 \\
        LSP/0 & 38.29 & 98.11 & 2.35 & 38.29 & 98.34 & 3.53 & 73.12 & 82.32 & 71.86 & 83.09 & 17.0 & {\bf 38.91} & 97.89 & 2.32 & {\bf 39.91} & {\bf 98.03} & 3.09 & 22.8 \\
        + 4-bit Quant. & 38.29 & 98.06 & {\bf 1.38} & 38.29 & 98.29 & {\bf 2.30} & 73.11 & 82.62 & 71.86 & 83.12 & 8.44 & {\bf 38.91} & 97.89 & {\bf 1.44} & {\bf 38.91} & {\bf 98.03} & {\bf 1.80} & 11.5 \\
        \hline
        \hline
        \end{tabular}
        }
\vspace{-1mm}
\end{table*}

\section{Experimental Studies}
\label{sect:eval}

\noindent
{\bf Datasets, metrics, and implementation details.} 
We use the MS MARCO Passage ranking dataset~\cite{Craswell2020OverviewOT} with 8.8 million English passages.  
We evaluate our results on the development (Dev) query set containing 6980 queries,
as well as the TREC deep learning (DL) 2019 and 2020 sets with 43 and 54 queries respectively.
We use the standard metrics of mean reciprocal rank (MRR@10) for the Dev set, 
and normalized discounted cumulative gain (nDCG@10) for DL19 and DL20.
We report recall at k (R@k) for all query sets.
For zero-shot retrieval performance, the BEIR~\cite{thakur2021beir} datasets are used,
and the size of each dataset varies from 3,633 to 5.4M documents. 
Following SP~\cite{2025SIGIR-Carlson-SP},
preserved recall ratio (recall budget) is the  recall percentage of approximate search divided by that of
safe search. 

Our implementation extends SP code~\cite{2025SIGIR-Carlson-SP} which is based on BMP using Rust, and compiled with -O3 optimization.
It will be released after the paper is published in a peer-reviewed venue. 
All timing results are collected after multiple runs
with a single thread on a Linux server using Intel i7-1260P with AXV2 SIMD support and 64GB memory. 
Before timing queries, all compressed posting lists and metadata for tested queries are pre-loaded into memory,
following the common practice.

\noindent
{\bf Setup.} 
We test our algorithm using two popular learned sparse retrieval algorithms, SPLADE++~\cite{Formal2021SPLADE, Formal_etal_SIGIR2022_splade++} 
and Efficient-SPLADE~\cite{lassance2022efficiency}.
The state-of-the-art baselines to be compared include 
SP~\cite{2025SIGIR-Carlson-SP}, BMP~\cite{mallia2024BMP}, and 
SeismicWave~\cite{2024CIKM-SeismicWave}, 
an extended version of Seismic~\cite{2024SIGIR-SparseApproximate}.
We do not compare ASC~\cite{2024EMNLP-ASC} because
SP is shown to outperform ASC~\cite{2025SIGIR-Carlson-SP} and ASC does not support compression.

\vspace{-2mm}
\subsection{Effectiveness on MS MARCO and BEIR}
{\bf In-domain search with MS MARCO passages on SPLADE}.
Table~\ref{tab:main} compares LSP/0 with other options based on two fixed configurations.
Config. 1 targets roughly 99\% relevance, while Config. 2 intends to be close to safe retrieval.
We use Config. 2 when searching the DL19/DL20 query sets.
Each index and config. are based on recommended parameters for each method. 

\noindent
{\bf MaxScore.}
We use safe search implemented in PISA~\cite{Turtle1995,mallia2019pisa}.

\noindent
{\bf BMP.} Config. 1 uses a query pruning value of $\beta=0.8$ (keep highest scoring 80\% of query terms) and Config. 2 is safe search.
Its indexes are built using $b=32,8$ for $k=10,1000$.

\noindent
{\bf SP.} Config. 1 uses $\mu=0.5,\eta=0.8$, and Config. 2 uses $\mu=0.5,\eta=1.0$. Indexes are built using $c=64$ and $b=16,8$ for $k=10,1000$.

\noindent
{\bf LSP/0.} For $k=10$, Configs. 1 and 2 use $\gamma=250,500$ and $k=1000$ uses $\gamma=1000,2000$.
For both $k$, $\beta=0.33,0.5$ are used for each config. respectively.
Indexes use $c=16$, and $b=16,4$ for $k=10,1000$.

\noindent
{\bf SeismicWave.} We use the best reported settings of 
\texttt{recall\_98}
and \texttt{recall\_99}~\cite{SeismicWave},
respectively for Configs. 1 and 2.
Both settings have a unique index.
Config. 1 uses 4 query terms, a heap-factor of 0.8, and unsorted blocks.
Config. 2 uses 6 query terms, a heap-factor of 0.8, and sorted blocks.
Note that these configurations were selected after an exhaustive grid search on the Dev set,
while all other methods could achieve better performance with additional tuning,
explored further in Table~\ref{tab:bestconfig-3versions}.
For $k=1000$, SeismicWave does not provide any configurations and
thus we use the indexing parameters of \texttt{recall\_99} and the same query processing parameters as $k=10$ while
doubling the posting length limit to 8000. That is  because the index used for $k=10$ could only achieve 97.6 recall.

Table~\ref{tab:main} show that without index compression, LSP/0 is generally the best method,
and it is up to 8.1x faster than BMP, 4.8x faster than SP, and up to 3.2x faster than SeismicWave.
Excluding SeismicWave for k=10, LSP/0 is always the fastest uncompressed method.
With compression and 4-bit block maximum weight quantization,
LSP/0 is fastest in all cases except SPLADE k=10.
SeismicWave is fastest for this setting (by 17-40\%),
which is likely because the recommended parameters come directly from a detailed grid search on the Dev set with SPLADE.
LSP/0 with 4-bit quantization does lose some quality compared to LSP/0
with 8-bit maximum weights.
However, it is still Pareto-optimal.
For example, 4-bit quantization loses 0.05 points of recall for SPLADE $k=1000$ Configuration 1 compared to 8-bit quantization. 
But, Config. 2 of 4-bit quantization gains 0.18 recall over 8-bit LSP/0 (Config. 1) in about the same time.
For DL19/DL20,
LSP/0 obtains 99\% of safe recall.
4-bit quantization again slightly affects retrieval quality, and its recall is slightly better than
8-bit for $k=1000$, while obtaining 99.9\% of safe recall for $k=10$.
The space for indexes for $k=10$ and 1000 will differ because the best indexing parameters were selected for each value of $k$.

\noindent
{\bf Zero-shot parameter robustness with LSR index and model variations.}
We examine if retrieval performance with the same parameters is robust to small variations of the index,
such as content updates and small model refinements.
For instance, a news index updated hourly can have large variations in the index's distribution, such as posting list length.
Here, it is not viable to run a full parameter grid search for each incremental change in index.
Optimal performance may not be required during such a change, but retrieval performance needs to remain competitive.

We use E-SPLADE to study the parameter robustness of LSP and baselines in adapting such a change
as it is from the same family as SPLADE but has a varation in index size (i.e. different lengths of posting lists). 
We keep the same configurations derived from SPLADE on MS MARCO and test them on E-SPLADE in a zero-shot setting.
For ESPLADE with $k=10$ and $\mu=0.5$,
SP failed to retrieve any documents for some queries because of the erroneous pruning discussed in Section~\ref{sect:design}.
While SeismicWave was faster than LSP/0 for $k=10$ on SPLADE,
when those same configurations are used on E-SPLADE,
compressed LSP/0 with quantization is up to 2.3x faster for $k=10$ while delivering similar relevance,
and up to 5.6x faster for $k=1000$.
Both BMP and LSP/0 achieve similar retrieval time and competitive relevance on both SPLADE and E-SPLADE using the same configurations.
Certainly, a model could achieve better performance with a grid search of parameters,
but this demonstates that LSP/0's and BMP's parameter choices are robust to model variations within the SPLADE family,
while SP fails and SeismicWave has degraded latency.

\noindent
{\bf Best configuration search with  LSP/0, LSP/1, and LSP/2 vs. SeismicWave on MS MARCO}. 
Table~\ref{tab:bestconfig-3versions} shows the performance of LSP/0, LSP/1, and LSP/2 when grid search of
the best configuration is allowed for MS MARCO passage top-10 and top-1000 retrieval.
We use $b=16,4$ for $k=10,1000$ respectively, $c=16$, and SIMDBP-256* with 4-bit quantization for all indexes.
For each column showing a fixed recall budget, 
LSP/1 generally achieves the best results for $k=1000$, but shows roughly similar performance to LSP/0 for $k=10$.
Both LSP/0 and LSP/1 are always better than LSP/2.

Table~\ref{tab:bestconfig-3versions} also shows a comparison of LSP to SeismicWave.
For each recall budget,
the row labeled ``Seismic-W.'' presents the recall of their best publicly provided configuration (e.g. \texttt{recall\_98}).
However, the best presented configurations never uses their KNN graph.
We run a grid-search to find the best configuration including the KNN graph for both $k=10$ and 1000.
These results show that without a KNN graph, SeismicWave is slower than LSP for $k=10$ and recall above 97\% of safe and faster from 93-97\%.
With a KNN graph, SeismicWave with a grid search is faster than LSP for $k=10$.
For $k=1000$, LSP is 38-87\% faster than Seismic-Wave, and the KNN graph hurts Seismic's performance.

\begin{table}[thbp]
    \centering
    \caption{Mean response time ($\mu s$) at a fixed recall budget using a grid search of parameters on MS MARCO Dev with SPLADE.
    All versions of LSP use a Fwd. document index.}
\vspace{-1em}
    \small
    \label{tab:bestconfig-3versions}
    \resizebox{0.7\columnwidth}{!}{
        \begin{tabular}{l|ccccc}
        \hline
        \hline
        {\bf Recall} & {\bf 93\%} & {\bf 95\%}& {\bf 97\%} & {\bf 98\%} & {\bf 99\%} \\
        {\bf Preserved} & MRT & MRT & MRT & MRT & MRT \\
        \hline
        & \multicolumn{5}{c}{$k$=10} \\
        \hline
        LSP/0 & 195 & 214 & 275 & 281 & 320  \\
        LSP/1 & 191 & 215 & 248 & 285 & 315  \\
        LSP/2 & 213 & 292 & 349 & 363 & 413  \\
        \hline
        Seismic-W. & 159 & 179 & 221 & 323 & 402 \\
        \  +$knn$ & {\bf 79} & {\bf 90} & {\bf 99} & {\bf 115} & {\bf 132} \\
        \hline
        \hline
        & \multicolumn{5}{c}{$k$=1000} \\
        \hline
        LSP/0 & 571 & 571 & 656 & {\bf 765} & 1030  \\
        LSP/1 & {\bf 516} & {\bf 562} & {\bf 647} & 780 & {\bf 961}  \\
        LSP/2 & 623 & 779 & 832 & 987 & 1330  \\
        \hline
        Seismic-W. & 713 & 876 & 955 & 1080 & 1800 \\
        \  +$knn$ & 2330 & 2330 & 2330 & 2360 & 2440 \\
        \hline
        \hline
    
        \end{tabular}
        }
\vspace{-1em}
\end{table}

\noindent
{\bf Zero-shot retrieval on the BEIR datasets}.
We seek to apply the recommended  $\gamma=250,\beta=0.33$ configuration (with 4-bit SIMDBP-256* compression) for LSP/0 when $k=10$
to see if lightweight pruning with a simple configuration
can achieve strong performance preserving 99\% of relevance while being fast for out-of-domain search without any training,
and how dataset size affects the efficiency of superblock pruning. 
While Seismic-Wave can outperform LSP/0 for k=10,
it requires an exhaustive grid-search of parameters to be effective (as shown on ESPLADE in Table~\ref{tab:main}),
and a cumbersome grid-search over indexing parameters which affect final retrieval performance. 
Moreove, the proposed optimizations of SeismicWave are not universally better: knn only benefits retrieval with $k=10$,
and sorting the blocks before evaluation is only used for some recall budgets. 
To compare SeismicWave in a true zero-shot manner, without a grid search of index or query time parameters,
we select their \texttt{recall\_99} configuration and index and retrieve according to parameters selected on MS MARCO.
We also compare versus SP and BMP (with compression).
We use the same indexing configuration for all datasets,
and choose $b=4$ because the largest datasets are about half the size of MS MARCO.
For $k=100$ and 1000, we increase $\gamma$ for LSP/0 to 500 and 1000 respectively,
following our results in Section~\ref{sect:analysis}, but do not adjust our query-pruning rate $\beta$.
The other methods use a fixed configuration for all value of k.

Table~\ref{tab:beir} shows the performance of BMP, SP, SeismicWave, and LSP/0 on the 13 BEIR datasets.
The row ``vs. LSP/0'' compares their performance to LSP/0 by calculating the average of the performance ratio between each method,
not the ratio of the average.
These results show that all methods maintained strong retrieval quality,
and that for all datasets except four, LSP/0 is fastest while maintaining competitive relevance.
On average, LSP/0 is twice as fast as SeismicWave, 5.7x faster than SP, and 9.3x faster than BMP.
It always has the smallest index size, and it maintains its relative latency advantage for $k=100$ and 1000.
This demonstrates that LSP/0's fixed configuration can effectively generalize to new datasets while maintaining
competitive relevance and latency.

\begin{table*}[thbp]
    \centering
    \caption{Zero-shot performance on 13 BEIR datasets.
             BMP uses $\beta=0.8, b=4$, SP uses $\mu=0.5$, $\eta=1.0, c=16, b=4$, LSP/0 uses $\gamma=250, \beta=0.33, c=16, b=4$, SIMDBP-256* with 4-bit weights, and Fwd. document index, Seismic uses \texttt{recall\_99} configuration.}
            \label{tab:beir}
    \vspace{-0.05cm}
    \resizebox{1.8\columnwidth}{!}{
    \begin{tabular}{l|c|c|ccr|ccr|ccr|ccr}

    \toprule   
    & \textbf{Corpus} & \textbf{Safe} & \multicolumn{3}{c|}{\bf BMP} & \multicolumn{3}{c|}{\bf SP} & \multicolumn{3}{c}{\bf SeismicWave} & \multicolumn{3}{c}{\bf LSP/0}\\
    \textbf{Dataset} & \textbf{Size} & nDCG & nDCG & MRT & GB & nDCG & MRT & GB & nDCG & MRT & GB & nDCG & MRT & GB \\

    \midrule
        \hline
        Arguana   & 8.7K & {\bf 52.0} & 51.1 & 0.414 & 0.070 & 48.7 & 0.704 & 0.096 &  49.5 & {\bf 0.089} & 0.188 & 50.3 & 0.222 & {\bf 0.023} \\
        C-FEVER   & 5.4M & 23.0 & 24.3 &  9.45 &  15.3 & {\bf 24.5} &  10.5 &  58.3 & 24.4 & {\bf 0.453} &  7.47 & 23.6 & 0.623 &  {\bf 6.14} \\
        DBPedia   & 4.6M & {\bf 43.7} & 43.3 &  6.34 &  19.8 & 43.6 &  2.02 &  48.3  & 43.2 & 0.381 &  5.63 & 42.9 & {\bf 0.330} & {\bf 4.79} \\
        FEVER     & 5.4M & 78.8 & 79.9 &  7.38 &  15.2 & 79.5 &  4.36 &  58.3 & {\bf 80.0} & 0.501 &  7.47 & 79.2 & {\bf 0.461} & {\bf 6.13} \\
        FiQA      &  57K & 34.7 & 35.5 & 0.639 & 0.405 & 35.5 & 0.598 & 0.666 & {\bf 35.6} & 0.841 & 0.713 & 35.1 & {\bf 0.186} & {\bf 0.093} \\
        Hotpot    & 5.2M & {\bf 68.7} & 68.6 &  9.62 &  21.4 & 68.5 &  3.73 &  54.0 & 68.3 & 0.442 &  5.73 & 67.1 & {\bf 0.422} & {\bf 5.04} \\
        NFCorpus  & 3.6K & 34.7 & 35.3 & 0.104 & 0.030 & 34.9 & 0.123 & 0.035 & {\bf 35.6} & {\bf 0.045} & 0.080 & 35.2 & 0.084 & {\bf 0.011} \\
        NQ        & 2.7M & 53.8 & 53.8 &  4.61 &  14.8 & {\bf 53.9} &  2.11 &  30.3 & 53.8 & 0.363 &  7.04 & 53.4 & {\bf 0.294} &  {\bf 3.57} \\
        Quora     & 520K & 83.4 & {\bf 83.5} & 0.945 &  1.27 & 83.2 & 0.521 &  4.51 & 83.4 & 0.305 & 0.693 & {\bf 83.5} & {\bf 0.193} & {\bf 0.294} \\
        SciDocs   &  25K & 15.9 & 15.8 & 0.444 & 0.213 & 15.9 & 0.499 & 0.306 & 15.8 & 0.239 & 0.413 & {\bf 16.2} & {\bf 0.175} & {\bf 0.052} \\
        SciFact   &   5K & 70.4 & 70.7 & 0.350 & 0.046 & 70.5 & 0.469 & 0.055 & 70.8 & {\bf 0.090} & 0.115 & 70.8 & 0.129 & {\bf 0.014} \\
        T-COVID   & 171K & 72.7 & 72.8 &  1.57 &  1.10 & 72.1 & 1.18 &  1.92 & 72.8 &  1.94 &  1.98 & {\bf 72.9} & {\bf 0.241} & {\bf 0.227} \\
        Touche    & 380K & 24.7 & {\bf 27.3} &  1.77 &  2.61 & {\bf 27.3} & 0.793 &  4.56 & 27.2 & 0.861 &  4.32 & {\bf 27.3} & {\bf 0.215} & {\bf 0.575} \\
         \midrule
        Average   & 1.9M & 50.5 & {\bf 50.9} & 3.22 & 7.10 & 50.6 & 2.12 & 20.1 & 50.8 & 0.504 & 3.22 & 50.6 & {\bf 0.275} & {\bf 2.07} \\
        vs. LSP/0 & -- & -0.7\% & +0.6\% & 9.3x & 3.7x & +0.2\% & 5.7x & 8.0x & +0.4\% & 2.0x & 5.0x & -- & -- & -- \\
         \midrule
         $k=100$ & 1.9M & 50.6 & {\bf 51.0} & 6.13 & 7.10 & 50.7 & 3.47 & 20.1 & 50.9 & 1.06 & 3.22 & 50.9 & {\bf 0.490} & {\bf 2.07} \\
         $k=1000$ & 1.9M & 50.6 & 50.9 & 13.9 & 7.10 & 50.7 & 8.75 & 20.1 & 50.9 & 2.52 & 3.22 & {\bf 51.0} & {\bf 1.31} & {\bf 2.07} \\
         \midrule

 \bottomrule

    \end{tabular}
    }
\end{table*}

\begin{table}[thbp]
    \centering
    \caption{Effect of $b$ and $\gamma$ on latency ($ms$) and recall (R@1k) of LSP/0 on MS MARCO Dev, SPLADE, $c=16$, Flat-Inv. }
    \label{tab:bsize_ablation}
    \small
    \resizebox{0.9\columnwidth}{!}{
        \begin{tabular}{l|cc|cc|cc|cc}
        \hline
        \hline
        & \multicolumn{2}{c|}{$\mathbf{\gamma=250}$} & \multicolumn{2}{c|}{$\mathbf{\gamma=500}$} & \multicolumn{2}{c|}{$\mathbf{\gamma=1000}$} & \multicolumn{2}{c}{$\mathbf{\gamma=2000}$} \\
        & MRT & R@1k & MRT & R@1k & MRT & R@1k & MRT & R@1k \\
        \hline
        {\bf b} & \multicolumn{8}{c}{$k$=1000} \\
        \hline
        4  & {\bf 2.12} & {\bf 95.9} & {\bf 2.59} & 97.4 & {\bf 3.28} & 98.1 & {\bf 4.31} & {\bf 98.3} \\
        8  & 2.55 & 95.5 & 3.31 & 97.3 & 4.40 & 98.0 & 5.60 & 98.2 \\ 
        16 & 3.94 & 95.0 & 5.53 & 97.0 & 7.35 & 97.9 & 9.48 & 98.2 \\ 
        32 & 6.68 & 94.0 & 9.87 & {\bf 97.8} & 13.8 & {\bf 98.2} & 17.8 & 98.2 \\ 
        64 & 11.3 & 92.9 & 21.2 & 95.8 & 25.7 & 97.7 & 33.6 & 98.2 \\ 
        \hline
        \hline
    
        \end{tabular}
        }
\end{table}

\noindent
{\bf Choice of block sizes}.
Table~\ref{tab:bsize_ablation} shows the top-1000 retrieval
latency of LSP/0 in milliseconds and recall when varying the block size from 4 to 64.
As $\gamma$ increases from 250 to 2000, bigger block sizes proportionally increase the retrieval latency.
Small block sizes tend to perform well because the superblock bound estimation is relatively more accurate.
Thus $b=8$ or 4 is a preferred block size for $k=1000$.
For $k=10$, the overhead of additional blocks outweigh the better bound estimation, and $b=16$ is best.

\noindent
{\bf A comparison of LSP/0, LSP/1, and LSP/2 with different parameters}.
Now  we show that sometimes LSP/1 can be slightly better than LSP/0 even though Table~\ref{tab:main}
uses LSP/0 with pre-determined configurations for simplicity without grid search.
Table~\ref{tab:SP3versions} compares top-1000 retrieval performance of
LSP/0, LSP/1, and LSP/2 under block size b=8 on MS MARCO Passages
with 4-bit quantization for block and superblock maximum weights. For LSP/2, we set $\eta=1$.
Notice that safe search achieves recall 98.36. 
Preserving that by 99\% means reaching recall $\ge$ 97.37.
With that as a target, LSP/1 with $\gamma=500$ and $\mu=0.20$ gives the shortest latency
3.61ms with recall 97.5.
Preserving safe recall by 99.5\% means reaching recall 97.87.
With that as a target, LSP/0 with $\gamma=1000$  gives the shortest latency 4.37ms with recall 98.
If we simply deploy LSP/0 with $\gamma=1000$, it can reach  a fairly high relevance budget
while its   speed is competitive to other options.

\begin{table}[thbp]
    \centering
    \caption{Comparison of LSP variants with SPLADE.
    $b$=8, $c$=16, 4-bit SIMDBP-256* compression, and Flat-Inv}
    \small
    \label{tab:SP3versions}
      \resizebox{\columnwidth}{!}{
        \begin{tabular}{l|cc|cc|cc|cc}
        \hline
        \hline
        & \multicolumn{2}{c}{$\mathbf{\gamma=250}$} & \multicolumn{2}{c}{$\mathbf{\gamma=500}$} & \multicolumn{2}{c}{$\mathbf{\gamma=1000}$} & \multicolumn{2}{c}{$\mathbf{\gamma=2000}$} \\
        & MRT & R@1k & MRT & R@1k & MRT & R@1k & MRT & R@1k \\
        \hline
        {\bf Method ($\mu$)} & \multicolumn{8}{c}{$k$=1000} \\
        \hline
        LSP/0        & 2.55 & 95.5 & 3.37 & 97.2 & 4.37 & 98.0 & 5.66 & 98.2 \\
        \hline
        LSP/1 (0.20) & 3.06 & 96.2 & 3.61 & 97.5 & 4.40 & 98.0 & 5.55 & 98.2 \\ 
        LSP/2 (0.20) & 3.71 & 96.2 & 4.39 & 97.5 & 5.11 & 98.0 & 6.06 & 98.2 \\
        \hline
        LSP/1 (0.25) & 3.78 & 96.9 & 4.13 & 97.7 & 4.69 & 98.1 & 5.60 & 98.2 \\ 
        LSP/2 (0.25) & 4.35 & 96.9 & 4.81 & 97.7 & 5.91 & 98.1 & 6.26 & 98.2 \\
        \hline
        LSP/1 (0.33) & 4.95 & 97.7 & 5.11 & 98.0 & 5.35 & 98.1 & 5.83 & 98.2 \\ 
        LSP/2 (0.33) & 5.51 & 97.7 & 6.00 & 98.0 & 6.18 & 98.1 & 6.52 & 98.2 \\
        \hline
        LSP/1 (0.50) & 7.11 & 98.3 & 7.07 & 98.3 & 7.12 & 98.3 & 7.19 & 98.3 \\
        LSP/2 (0.50) & 7.80 & 98.3 & 7.56 & 98.3 & 8.01 & 98.3 & 7.63 & 98.3 \\
        \hline
        LSP/1 (0.75) & 8.82 & 98.3 & 8.86 & 98.3 & 8.85 & 98.3 & 8.95 & 98.3 \\
        LSP/2 (0.75) & 9.04 & 98.3 & 9.23 & 98.3 & 8.93 & 98.3 & 9.13 & 98.3 \\
        \hline
        \hline
    
        \end{tabular}
        }
\end{table}

\subsection{Compression evaluation}
{\bf Compression Options and Index Size}.
\label{sect:eval_comp}
Table~\ref{tab:index_size} lists the in-memory storage size of document-level index when LSP uses 
original BMP-Inv, 
its optimized Compact-Inv, the proposed Flat-Inv, and a forward index (Fwd.).
Separately, it lists the total storage for block and superblock maximum weights when LSP  uses
four options: uncompressed (BMP-Dense),  BMP-Sparse, and SIMDBP-256* with 8-bit maximum  weights,
and  SIMDBP-256* with 4-bit maximum  weights.
The total amount of in-memory storage required is the sum of space for block/superblock maximum weights, and 
for document-level index.
For example, with $b=8$, our proposed method with  SIMDBP-256* and 4-bit block/superblock quantization would 
require $3.6 + 3.5 \approx 7.1$ GB in-memory.

For the document-level index within each block, the proposed flat compact structure constantly delivers the lowest storage size of all inverted indexes 
with up to 6.2x and 5.3x  space reduction compared to the BMP method when $b=2$ and $b=8$, respectively.
Seismic's forward index uses less storage when $b\leq 16$, and offers up to a 9x space reduction compared to BMP-Inv when $b=2$.
For block/superblock level weights,
SIMDBP-256* with 4-bit quantization has the smallest index size for all configurations. 
\begin{table}[htbp]
    \small
        \centering
        \caption{Comparison of compression methods for LSP: MS MARCO in-memory index size (GB) with SPLADE.  $c$=16} 
    \label{tab:index_size}
    \resizebox{0.9\columnwidth}{!}{
        \begin{tabular}{l|cccccccc}
        \hline
        Block Size & 2 & 4 & 8 & 16 & 32 & 64 & 128 & 256 \\
        \hline
        \multicolumn{9}{l}{ {\bf Document index } } \\
        BMP-Inv & 28 & 23 & 19 & 15 & 12 & 9.6 & 7.6 & 6.1 \\ 
        Compact-Inv  & 22 & 18 & 14 & 12 & 9.5 & 7.7 & 6.3 & 5.1 \\
        Flat-Inv  & 4.5 & 4.0 & 3.6 & 3.2 & {\bf  2.9} & {\bf 2.6} &{\bf  2.4} &{\bf  2.4} \\
        Fwd. & {\bf 3.1} & {\bf 3.1} & {\bf 3.1} & {\bf 3.1} & 3.1 & 3.1 & 3.1 & 3.1 \\
        \hline
        \multicolumn{9}{l}{ {\bf Superblock/block maximum weights}} \\ 
        BMP-Dense & 125 & 63 & 32 & 16 & 8.0 & 4.0 & 2.0 & 1.0 \\
        BMP-Sparse & 15 & 9.8 & 7.6 & 5.6 & 4.2 & 3.0 & 2.2 & 1.6 \\
        SIMDBP-256* & 21 & 14 & 9.4 & 6.0 & 3.7 & 2.3 & 1.3 & 0.75 \\
         + 4-bit Quant. &{\bf  7.4} &{\bf  5.3} &{\bf  3.5} & {\bf 2.4} & {\bf 1.5} & {\bf 0.95} & {\bf 0.59} & {\bf 0.33} \\
        \hline

    \end{tabular}
    }
\end{table}

\noindent
{\bf Compression ablation study}. 
Table~\ref{tab:index_speed} examines retrieval latency and  overall space of combining
key compression techniques presented in Section~\ref{sect:compress} under two block sizes and three recall budgets. 
In a row-wise top-down manner, we start with an uncompressed index (i.e. BMP-Dense), and then gradually apply selected compression methods.
For superblock/block maximum weight compression, we first apply BMP-Sparse or SIMDBP-256*. For SIMDBP-256*, we further
apply 4-bit quantization. 
After selecting SIMDBP-256* with 4-bit quantization,
we additionally apply compression to the document index using Compact-Inv, Flat-Inv, or Seismic-Fwd.
Times reported for 4-bit quantization are not equivalent to 8-bit safe and are distinguished with asterisks. 
These methods are about 99.8\% of 8-bit safe.
The results show that a forward index with SIMDBP-256* and 4-bit maximum weight quantization
has a reasonably fast speed while the smallest space (6.8GB) with $b=8$.
For $b=128$, using Compact-Inv for document inverted index is faster than both Flat-Inv and Seismic-Fwd but costs much more space.
Flat-Inv uses slightly less space than Seismic-Fwd for $b=128$.

\begin{table}[htbp]
    \small
        \centering
        \caption{MS MARCO search time and index size with SPLADE when adding different compression techniques for LSP/1.
}
    \label{tab:index_speed}
    \resizebox{\columnwidth}{!}{
        \begin{tabular}{l|ccc|ccc|cc}
        \hline
        	& \multicolumn{3}{|c|}{Latency (ms), b=8} &\multicolumn{3}{|c}{Latency (ms), b=128} & \multicolumn{2}{|c}{Space (GB)}\\
        Recall Budget & 97\% & 99\% & Safe & 97\% & 99\% & Safe & b=8 & 128 \\
        \hline
        Uncompressed & {\bf 0.51} & 0.84 & {\bf 3.3} & 12.0 & 13.3 & 14.2 & 51 & 9.6 \\ 
        \hline
        \multicolumn{8}{l}{ {\bf Add  superblock/block maximum weight compression} } \\
        BMP-Sparse   & 2.7 & 18 & 250 & 73 & 106 & 240 & 26 & 9.8 \\
        SIMDBP-256* & 0.83 & 1.7 & 6.1 & 12.2 & 13.5 & 14.6 & 28 & 8.9 \\ 
        \  +4-bit Quant. & 0.52 & 0.82 & 4.7* &  7.8 &  8.3 & 8.9* & 22 & 8.2 \\
        \hline
        \multicolumn{8}{l}{ {\bf Add document index compression} } \\
        Compact-Inv & 0.59 & 0.95 & 5.4* & {\bf 7.5} & {\bf 7.9}& {\bf 8.6*}& 18 & 6.9 \\ 
        Flat-Inv & 0.56 & 0.85 & 4.7* & 9.4 & 9.9 & 10.7* & 7.3 & {\bf 3.2} \\ 
        Fwd. & 0.52 & {\bf 0.77} & 4.8* & 9.9 & 10.6 & 11.8* & {\bf 6.8} & 3.7 \\
        \hline

    \end{tabular}
    }
\end{table}

\noindent
{\bf Inverted vs. forward index per document block}.
Table~\ref{tab:fwd_comparison} compares the latency of Flat-Inv and Fwd.
across different block sizes for both safe and approximate retrieval.
The forward index achieves lower latency at small block sizes ($b \leq 64$),
but similar or higher latency than Flat-Inv at larger block sizes ($b \geq 96$).
Fwd. stores separate arrays for term ids and weights,
requiring only two random memory accesses per block.
However, it must fetch all terms in each document, regardless of which terms appear in the query.
The inverted index requires one random memory access per query term ($\sim$43 for SPLADE++ on the Dev set),
but only fetches postings for terms present in the query.
At small block sizes, the random memory accesses of the inverted index cause it to be slower,
while at larger block sizes the total memory fetched per block becomes a bottleneck,
favoring the inverted index.
\begin{table}[htbp]
    \small
        \centering
        \caption{Latency comparison (ms) of Flat-Inv vs. Fwd. document indexes on MS MARCO with SPLADE for LSP/0 under different block sizes. $c$=16, $k$=1000, SIMDBP-256*.}
    \label{tab:fwd_comparison}
    \resizebox{0.8\columnwidth}{!}{
        \begin{tabular}{l|cccccc}
        \hline
        & \multicolumn{6}{c}{Block Size ($b$)} \\
        Index Type & 8 & 16 & 32 & 64 & 96 & 128 \\
        \hline
        \multicolumn{7}{l}{ {\bf 99\% recall budget}} \\
        Flat-Inv & 4.29 & 9.00 & 17.0 & 31.3 & 44.0 & {\bf 55.6} \\
        Fwd. & {\bf 2.43} & {\bf 5.30} & {\bf 11.4} & {\bf 25.8} & {\bf 43.1} & 60.5 \\
        \hline
        \multicolumn{7}{l}{ {\bf Safe retrieval}} \\
        Flat-Inv & 11.6 & 17.0 & 31.3 & 56.4 & 76.5 & {\bf 91.6} \\
        Fwd. & {\bf 9.65} & {\bf 11.5} & {\bf 22.1} & {\bf 48.0} & {\bf 76.3} & 103 \\
        \hline
    \end{tabular}
    }
\end{table}

\vspace{-3mm}
\section{Conclusion}

We introduced the safe inclusion of top-$\gamma$ superblocks 
in superblock- and block-based retrieval to avoid erroneous index skipping and to simplify pruning with less overhead.
The main takeaways are:
\begin{itemize}[leftmargin=*]
\item {\bf Out-of-domain zero-shot on BEIR.} 
LSP/0 with a simple configuration is 2x faster than SeismicWave, 5.7x faster than  SP, and 9.3x faster than BMP 
with negligible quality differences,
while requiring 5x, 8x and 4x less memory respectively.
Thus for out-of-domain retrieval, the main advantage of LSP/0
over SeismicWave  is its simplicity without configuration grid search and avoiding the 
need  of a KNN  graph which is complex to maintain with all-pair document-level comparison during search data update. 
\item {\bf In-domain search with MS MARCO.}
LSP/0 is 1.8x to 12x faster than BMP while using ~3.3x less memory,
and up-to 17x faster than SP while using 1/6th the memory.
LSP/0 with a simple configuration 
is up-to 3.2x faster than SeismicWave
for $k=1000$ on SPLADE,
but it is about 30\% slower for $k=10$ on SPLADE.
When a grid search of all parameters is possible with SPLADE,
for $k=1000$, LSP still outperforms SeismicWave by 55\% to 78\%.
For $k=10$,
LSP/0 is about 20\% faster than SeismicWave without a KNN graph above 98\% preserved recall,
while about 20\% slower for 93-97\% preserved recall.
With a KNN graph, SeismicWave is about 2x as fast as LSP for $k=10$. 

\item {\bf Parameter adaptivity.}
LSP/0 with our proposed configuration derived from SPLADE
is able to be applied to E-SPLADE on MS MARCO successfully.
Under the same setting, BMP succeeded, SP failed from erroneous pruning,
 and SeismicWave's latency roughly doubled.
 LSP/0 was the fastest of the three on E-SPLADE.

\item {\bf LSP/0 vs LSP/1 vs LSP/2.}
LSP/1 can outperform LSP/0 in some cases when using a parameter grid search, 
however LSP/0 is easy-to-use with highly competitive performance both with and without compression.
Both LSP/0 and LSP/1 outperform LSP/2, which shows that superblock average weights are not needed when 
using a large number of blocks and a top-$\gamma$ guarantee.

\item {\bf Recommended LSP configuration.}
LSP/0 with $\gamma=250$ or 500 is recommended when $k=10$, and  $\gamma=1000$ or 2000 when $k=1000$,
with query pruning set to $\beta\approx 0.33$.
We recommend SIMDBP-256* compression is used with 4-bit quantization,
and a forward document index for small values of $b$.
LSP performs best when the block size is small (e.g. $4\leq b \leq 16$)
and the superblock size is within the tested range:  $b\times c \leq 256$.
More work is needed to test different superblock sizes.
This configuration provides competitive performance, albeit not the best,
and generalizes well across
multiple datasets and  models, which
greatly simplifies its use for zero-shot out-of-domain search applications.  
\end{itemize}

Overall, LSP is  a strong  alternative for efficient retrieval.
Our future work is to enhance LSP with static pruning like Seismic,
and explore effective ways to incorporate a proximity graph.

\balance
\bibliographystyle{ACM-Reference-Format}
\normalsize
\bibliography{bib/2025.bib,bib/2024refer.bib,bib/thres.bib,bib/2022extra.bib,bib/2022refer.bib,bib/jinjin_thesis.bib,bib/mise.bib,bib/ranking.bib,bib/reference.bib,bib/url.bib,bib/distill.bib}

\end{document}